\documentclass[aps,prl,amssymb,twocolumn]{revtex4-1}
\usepackage{graphicx}
\usepackage{dcolumn}
\usepackage[T2A]{fontenc}
\usepackage[cp1251]{inputenc}

\begin{document}

\title{Localized superconductive pairs}

\author{V. F. Gantmakher}

\affiliation{\mbox{Institute of Solid State Physics RAS, 142432 Chernogolovka,
Moscow region, Russia}\\ E-mail:\rm\ gantm@issp.ac.ru}

\begin{abstract}
Different physical phenomena are discussed which should help to comprehend and
interpret the concept of localized superconductive pairs; these include
behavior of highly resistive granular materials with superconducting grains,
parity effect and the Berezinskii--Kosterlitz--Thouless transition.
Experimental arguments in support of localized pairs existence are presented and
conditions which promote their appearance are analyzed.

\vspace{1cm} \normalsize
\noindent 1.Introduction\\
 2. Granular systems

\hspace{3mm} 2a. Pseudolocalization of Cooper pairs in granular metals

\hspace{3mm} 2b. Parity effect in small grains\\
3. Delocalized electron pairs in dissipative environment

\hspace{3mm} 3a. Berezinskii--Kosterlitz--Thouless transition

\hspace{3mm} 3b. Superconductivity as a Bose-condensation process  \\
4. Experimental verification of existence of localized pairs

\hspace{3mm} 4a. Negative magnetoresistance

\hspace{3mm} 4b. Binding energy of localized electron pairs --- superconductive pseudogap

\hspace{3mm} 4c. Size of localized pairs \\
5. Additional factors that promote pairs localization

\hspace{3mm} 5a. Proximity of metal--insulator transition, either real or
virtual

\hspace{3mm} 5b. "Chemical predisposition"\ to pair localization\\
Acknowledgement\\
References
\end{abstract}

\maketitle

The idea that the superconducting interaction may not only maintain a
dissipationless electric current through the metal but also assist localization
and establish insulating state seems paradoxical and hardly opens the way. This
idea rests upon the model of localized superconductive pairs. This paper
contains its comprehensive analysis. The model concretizes idea about
Bose-insulator or Bose-glass introduced for the first time when Bose-Einstein
condensation of charged Bose-gas in the field of charged impurities was studied
in \cite{Gold} and unraveled in detail in \cite{FF}, where the properties of a
system of bosons with weak repulsion was investigated; the bosons were arranged
at sites in the lattice with a finite probability of hopping between the sites.

The paper starts with definitions and general notations. Then short description of zero-dimensional superconductivity of separate grains and behavior of assemble of such grains is presented followed by the parity effect in an isolated small grain. Thus, one logic path for understanding of the localized pairs phenomenon is proposed --- from grains toward point defects.

Another logic path also can be paved: from superconductor to dissipative medium
with equilibrium concentration of incoherent Cooper pairs, then to insulator with localized pairs. The first phenomenon situated along this path is the Berezinskii--Kosterlitz--Thouless transition: nonzero concentration of Cooper pairs exists on both sides of transition of a two-dimensional superconductor into non-dissipative state. Next, besides the main Bardeen--Cooper--Schrieffer model of superconductivity (BCS) another model exists which assumes superconductor
transition to be a sort of Bose-Einstein condensation (BEC). It seems
reasonable to use this model or some its modification for systems with small
electron concentration (see, for instance, review \cite{Loktev}). BEC model
also operates with electron pairs (bosons) and they exist on both sides of BEC transition. With these two examples in mind, it becomes much easier. If we accept existence of uncorrelated electron pairs in dissipative medium, it is natural to assume that they can be localized.

These excursus set forth in the second and third sections should help to accept experimental evidences that localized pairs do exist presented in the forth section: negative magnetoresistance in the vicinity of the superconductor--insulator transition on its insulating side, first measurements of the density of states near the Fermi level in this insulator and the first successful attempt to measure the effective volume occupied by a localized pair.

In the last section, factors which are important for pair localization are
discussed: proximity to the insulator--normal metal transition and specific
correlations in the random potential, in particular, those determined by the
chemical content of material.

The paper rests upon recent review \cite{GD2} on the superconductor--insulator transitions.

\section{1. Introduction}

Electron is assumed localized on defect or center if its stationary wave
function decays at large distance $r$ exponentially
\begin{equation}\label{aB}
  \psi=\psi_0\exp(-r/\Lambda),\qquad r\gg\Lambda,
\end{equation}
with $\Lambda$ called the localization length.

The electron localization length in an isolated defect is usually called the Bohr radius $a_B$. Since localization takes place against the background of random
potential, electron energy at various centers slightly differs. However, at
finite temperatures $T\neq0$, the electron can hop from one center to another,
with the energy being conserved due to some accompanied processes, for
instance, emission or absorption of phonons. Such electron hops lead to hopping
conductivity. The low temperature hopping conductivity $\sigma_h$ usually is
described by formula
\begin{equation}\label{hopping}
\sigma_h=\sigma_{h0}\exp[-(T_1/T)^{1/\nu}],\qquad\nu=1,2,3 \mbox{ or } 4;
\end{equation}
specific value of $\nu$ depends on the type of hopping conductivity, density of
states $g(\varepsilon)$ near the Fermi level $\varepsilon_F$ and
dimensionality.

When exponential tails of wave functions of electrons localized at different
centers overlap, the localization length $\Lambda$ increases and becomes larger
then $a_B$. In particular, this happens while approaching the insulator--metal
transition; straight at the transition $\Lambda$ becomes infinity, so that
while approaching the transition it spans the interval
\begin{equation}\label{Lambda}
  a_B\leqslant\Lambda\leqslant\infty.
\end{equation}

Generally speaking, interaction between localized electrons comes not only from
overlapping of their wave functions. In particular, one may imagine that
between two such electrons additional superconducting interaction arises, i.e.
they exchange virtual phonons. As a result, energies of both electrons becomes
amount $\Delta_L$ less. For phonon exchange to take place, the level spacing of
each electron, determined by its localization volume, should be less then the
phonon energy. If such indirect interaction takes place indeed, wave function
of the localized pair may be introduced. At large distances, it also decays
exponentially
\begin{equation}\label{a2B}
  \psi_2=\psi_{20}\exp(-r/\Lambda_2),\qquad r\gg\Lambda_2,
\end{equation}
with typical length $\Lambda_2$. In the vicinity of insulator--superconductor
transition, relation for $\Lambda_2$, similar to (\ref{Lambda}) may be written:
\begin{equation}\label{Lambda_2}
  a_{2B}\leqslant\Lambda_2\leqslant\infty.
\end{equation}

Hopping conductivity in the insulator with paired electrons is also possible.
For this, one of two events should occur. Either two electrons hop
simultaneously to new centers remaining interconnected; this process is still
almost not studied. It may be implemented in the close vicinity of the
superconductor--insulator transition --- we shall return to this possibility
below, at the end of the section 4. Or an electron "pays off"\ its partner
leaving him additional energy $\Delta_L$; contribution of such hops with
dispairing to the conductivity has additional small factor
\begin{equation}\label{sigma_2}
  \Delta\sigma_h^{(2)}=\Delta\sigma_h\exp(-\Delta_L/T).
\end{equation}
Anyway, one should expect that activation conductivity of insulator I$_2$ with
paired carriers is less then of an usual insulator.

The aim of this paper is to inquire into reality of such model of localized
pairs and to study experimental facts that may be interpreted as proof of existence of the pairs.

\section{2. Granular systems}

{\bf 2a. Pseudolocalization of Cooper pairs in granular metals}

Let us examine material which consists of grains of superconductive material in
insulating matrix. Let the mean size $b$ of the grains ensures the relation
\begin{equation}\label{b>}
\Delta/\delta\varepsilon=\Delta g_Fb^3\gg1,
\end{equation}
with $\Delta$ being the superconducting gap; $\delta\varepsilon=(g_Fb^3)^{-1}$
is electron level spacing determined by the grain size and $g_F$ is the density of states at the Fermi level in the bulk metal in the normal state. It is just
existence of the gap  in comparatively dense system of electron levels that
allows to call the grain superconducting.

Expression in the left side of inequality (\ref{b>}) can be interpreted as the
number of Cooper pairs in the grain:
\begin{equation}\label{n2b3}
(g_F\Delta)b^3\gg1.
\end{equation}

Superconducting state of a grain is a joint state of all Cooper pairs. It is
characterized, as in bulk superconductor, by the complex order parameter
\begin{equation}\label{OrderParameter}
  \Phi{(\bf r)}=|\Phi|\exp(i\varphi(\bf r)),
\end{equation}
in which the value of the gap $\Delta$ in the spectrum is used as modulus,
$|\Phi|=\Delta$, while the phase $\varphi(\bf r)$ characterizes coherence of
Cooper pairs. If there is no current in the superconductor, then $\varphi(\bf
r)=\mathrm{const}$.

The charge transfer from one grain to another is possible only by tunneling.
If this transfer is accomplished by Josephson currents of Cooper pairs, then
the phase values of the order parameter in all grains are correlated and
macroscopic superconductive state is set in the material. However, Josephson
currents may be damped, for instance, by too large tunnel resistance
$\rho\gg\hbar/e^2$ between the grains. Then the charge transfer between the
grains is maintained by one-particle excitations. Their concentration $n_1$ is
exponentially small because of the superconducting gap $\Delta$:
\begin{equation}\label{n1}
n_1\propto\exp(-\Delta/T).
\end{equation}
Under these conditions, the superconductivity sets in independently in each
grain. One might say that pairs are localized inside their grains with
localization radius $\Lambda_2$ equal to the grain size $b$: $$\Lambda_2=b.$$
If the grain size $b$ is macroscopic and relation (\ref{n2b3}) is satisfied,
such use of the word "localization"\ looks doubtful. However, it is useful from
two points of view.

\begin{figure}
\includegraphics{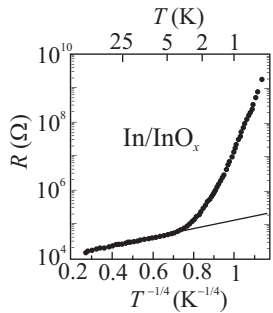}
\caption{Temperature dependence of the resistance of granular film with In
grains parted by insulating layers of the oxide \cite{Kim}. The function $R(T)$
demonstrates paradoxical \emph{increase} of the resistance below the
temperature of the superconducting transition in In grains. For bulk pure In, $T_c=3.41$\,K, $T_c^{-1/4}=0.735$\,K$^{-1/4}$ } \label{Kim}
\end{figure}

First, it shows that the idea about localized superconductive pairs is not as
insane as it may appear. Second, it induces to examine closely transport
properties of high resistive granular superconductors. These properties are
demonstrated by Fig \,\ref{Kim}. It presents the temperature dependence of the
resistance for the granular system In/InO. The normal resistance $R\equiv R_n$
increases according to the Mott law, as in usual insulator, until the
temperature $T$ remains above the temperature $T_c$ of the superconducting
transition of indium. The superconducting transition undeniably affects the
$R(T)$ dependence but the resistance $R\equiv R_{sc}$ starts to increase faster
instead of lessening.

Qualitatively, this effect is quite understandable. At low temperatures,
one-particle excitations which can tunnel between the grains die out in accordance
with formula (\ref{n1}). This process superimposed on the others, usual ones,
introduces additional factor $$R_{sc}=R_n\exp(\Delta/T)$$ into the expression
for $R_{sc}$. When magnetic field destroys the superconductive gap inside the
grains and makes them normal, the number of one-particle excitations at the
Fermi level increases and the resistance returns to the value $R_n$. This means
the stronger negative magnetoresistance the lower is the temperature
\begin{equation}\label{OMR}
  R(B,T)/R(0,T)\approx\exp(-\Delta/T),
  \qquad B>B_c,
\end{equation}
with $B_c$ being the critical field.

One can apply similar scheme to hypothetical localized pairs in a quasi homogeneous material, after changing the word "tunneling"\ with "hopping"\ and assuming that magnetic field destroys the superconductive correlations in the pairs. As the localized pairs should inherit singlet state from Cooper pairs so that two electron spins in a pair are of opposite directions, the magnetic field tending align the spins decreases the energy gain from pairing and the field strong enough blanks it.
Of course, the assumption of such response of localized pairs to the field must be checked experimentally. Anyway, \emph{experiments with granular superconductors indicate the direction for searching of electron pair localization. If there are pairs localized on defects one can expect negative magnetoresistance in strong magnetic field at the expense of destruction of these pairs.}

\vspace{2mm}{\bf 2b. Parity effect in small grains.} Due to Coulomb blockade, the parity effect was observed \cite{Tinkham,Devoret} in superconductive grains with size
satisfying inequality (\ref{b>}). Adding of additional odd electron to the system increased the full electron energy $E_N$ more than adding of the next even electron. The difference is $2\Delta_p$ with
\begin{equation}\label{Delta_p}
 \Delta_p=E_{2l+1}-\frac{1}{2}(E_{2l}+E_{2l+2})
\end{equation}
being the binding energy per one electron. In large grains $\Delta_p=\Delta$.

When the size of the grain  decreases, then condition (\ref{b>}) breaks and the superconductivity must disappear. This happens at  $b<b_1$, with $b_1$ being determined by comparison of superconducting gap $\Delta$ and spacing $\delta\varepsilon$
\begin{equation}\label{g0}
  \delta\varepsilon=(g_Fb_1^3)^{-1}=\Delta,\qquad b_1=(g_F\Delta)^{-1/3}.
\end{equation}

However, the phonon-mediated superconducting interaction remains and it still leads to effective interelectron attraction. According to theoretical findings \cite{Matveev}, the parity effect also remains. The binding energy $\Delta_p$ in small grains
\begin{equation}\label{smallGrains}
 b\ll b_1,\qquad\mbox{i.e.}\qquad\delta\varepsilon\gg\Delta,
\end{equation}
becomes small correction relative to $\delta\varepsilon$. However, it is not small comparing to gap $\Delta$ in the bulk:
\begin{equation}\label{Delta_p-small}
 \Delta_p=\frac{\delta\varepsilon}{2\ln(\delta\varepsilon/\Delta)}>\Delta.
\end{equation}

Findings \cite{Matveev} are valid until spacing remains less then Debye phonon energy
$\hbar\omega_D$. This means that there exists some interval for
\emph{the size of isolated grain of superconductive metal }
\begin{equation}\label{b1b2}
\begin{array}{c}
 b_2\ll b\ll b_1,\\ b_2=(g_F\hbar\omega_D)^{-1/3}, \qquad
 b_2/b_1\approx(\Delta/\hbar\omega_D)^{1/3},
\end{array}
\end{equation}
where \emph{superconducting interaction may lead to pairing of electrons localized inside the non superconducting grain on the size $b$.}

\section{3. Delocalized electron pairs in dissipative environment}

According to the classical BCS theory of superconductivity, equilibrium finite concentration of Cooper pairs in the bulk appears at temperature $T_c$ simultaneously with dissipationless state; we assume this state to be indeed superconductive.The superconductivity is usually destroyed by nulling the module of the order parameter $|\Phi|\equiv\Delta$; this can happen, for instance, with temperature growth up to
$T_c$ or field growth up to $B_c$. In this section, we shall study possibility for two following events to occur \emph{separately}: destruction of superconductivity and nulling of $|\Phi|$, i.e. possibility for dissipation to appear at nonzero equilibrium concentration of electron pairs. We shall consider two cases.

 \vspace{2mm}{\bf 3а. Berezinskii--Kosterlitz--Thouless transition.}

Two-dimensional superconducting systems have a specific feature: below the temperature $T_{c0}$ of bulk superconducting transition, finite concentration of Cooper pairs coexists in some temperature interval with gas of fluctuations in the form of spontaneously generated magnetic vortices. Each vortex contains magnetic flux quantum
\begin{equation}\label{Phi0}        
\Phi_0=2\pi\hbar c/2e.
\end{equation}

Vortices are created paired, with opposite directions of the field along the axis (pairs vortex-antivortex). They have finite life-time and annihilate while collisions. In zero magnetic field, concentrations of vortices with opposite sign are equal:
$N_+=N_-$; they are determined by dynamic equilibrium between the processes of spontaneous generation and annihilation. Path-tracing around a vortex changes the phase of the order parameter by $2\pi$; hence free motion of the vortices results in phase fluctuations of the order parameter. High enough fluctuation amplitude destroys coherence of the electron system. However, module of the order parameter remains finite in the most part of the volume; it is zero only inside a vortex, near its axis.

With lowering the temperature, Berezinskii--Kosterlitz--Thouless (BKT) transition takes place \cite{Berez, KostThoul} at some $T_c<T_{c0}$: generation of vortex pairs stops, vortex concentration slumps and becomes at $T<T_{c}$ exponentially small, together with with dissipation. Hence, Cooper pairs coexist in two-dimensional superconductors with vortices in the temperature interval
\begin{equation}\label{BKT}
 T_c<T<T_{c0}.
\end{equation}
Existence of Cooper pairs in this interval and their partial coherence decrease the dissipation but do not annihilate it.

From mathematical point of view, existence of dissipationless state means that correlator
\begin{equation}\label{Correl}
  G({\bf r})=\langle\Phi({\bf r})\Phi(0)\rangle\rightarrow G_0\neq0\quad
  \mbox{при}\quad|{\bf r}|\rightarrow\infty
\end{equation}
remains finite at large distances (angle brackets mean averaging over quantum state of the system). In the temperature interval (\ref{BKT}), correlator (\ref{Correl}) tends to zero with increase of $\bf r$ exponentially. At temperatures $T<T_c$, it decreases in accordance with a power law, i.e. it tends to zero all the same but comparatively slowly. Therefore, in two-dimensional superconductors, we have dissipative state in the temperature interval (\ref{BKT}) and almost coherent state at $T<T_c$; strictly coherent state in a two-dimensional superconductor is settled only at $T=0$.

\begin{figure}[h]
\includegraphics{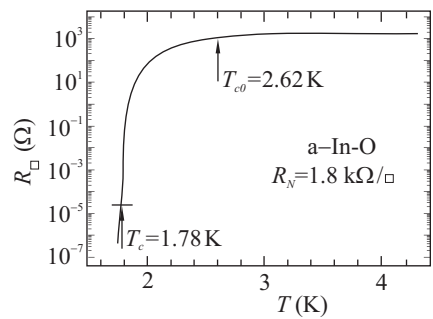}
\caption{Temperature $T_{c0}$ when equilibrium concentration of Cooper pairs appears and temperature $T_c$ when coherent state is established in the film In-O with thickness 100\,\AA, tied to the curve of resistive transition, \cite{Hebard0}. } \label{Hebard0}
\end{figure}

Fig.~\ref{Hebard0} demonstrates an example of layout of temperatures $T_{c0}$ and $T_c$ along the curve of the resistive transition. It was determined in
\cite{Hebard0} for superconductive transition in amorphous films In-O based on rigorous application of theoretical calculations to experimental data.
According to the figure, there is almost 50\% difference between the temperatures
$T_{c0}$ and $T_c$; $R(T_{c0})\approx0.5R_N$, and $R(T_c)$ is by several orders of magnitude less then $R_N$ ($R_N$ is the resistance of the film in the normal state).

 \vspace{2mm}{\bf 3b. Superconductivity as a Bose-condensation process.}
The BCS theory assumes that the coherence length (Cooper pair size)
$\zeta\sim\hbar v_F/\Delta\sim10^{-4}$ cm is much larger than mean distance between pairs $s\sim (g_0\Delta)^{-1/3}\sim10^{-6}$ cm ($g_0$ is normal metal density of states at the Fermi level):
\begin{equation}\label{1}
  \zeta\gg s.
\end{equation}
As a matter of fact, it is collective state of all electrons that is presented as aggregation of Cooper pairs.

However, superconductivity sometimes appears in systems with electron concentration perceptibly lower than in standard metals, for instance, in
SrTiO$_3$ with the electron concentration of the order
$n\sim10^{19}\,$cm$^{-3}$ \cite{SrTiO}. In addition, parameter $\zeta$  in type-II superconductors can be less than 100\,\AA. Hence the inequality (\ref{1})
required for the BCS model to be used can prove to be violated. Materials with $\zeta\lesssim s$ are referred to as exotic superconductors.

The existence of exotic superconductors, for which inequality (\ref{1}) is violated, forced to turn to another model of superconductivity, the Bose--Einstein condensation of the gas of electron pairs considered as bosons with charge $2e$ \cite{Schafroth} and to investigate crossover from the BCS to the BEC model (see, e.g., the review \cite{Loktev}).

The BEC model assumes another way of destruction of superconductive state: phase fluctuations of the order parameter vanish the correlator (\ref{Correl}), while modulus of the order parameter remains finite, \cite{VarLark}. Finite modulus of the order parameter at the transition means finite concentration of bound electron pairs, i.e. boson concentration at the transition point does not vanish. Realization of such scenario is favored by the fact that the superconductors with a low electron density are characterized by a weaker shielding and a comparatively small "rigidity"\ relative to phase changes, thus raising the role of phase fluctuations, \cite{EmKiv1,EmKiv2}.

Though BEC model assumes existence of bosons (electron pairs) on both sides of the superconductive transition, it does not discuss how pairs appear above $T_c$.
One can suppose, for instance, that pairs appear regardless of each other due to Cooper interaction. Their relative concentration is determined by temperature; at high temperature $T\gg\Delta$ it becomes negligibly small.

After one acknowledges possibility for uncoherent electron pairs in dissipative environment to exist, the question about possible localization of these pairs becomes quite natural with answer depending on the level of disorder. Equilibrium concentration of electron pairs means that a gap or, at least, minimum in the one-particle density of states exist at the Fermi level of the electron system. \emph{Measurements of the density of states on both sides of the superconductive transition and search for its minimum at the Fermi level on the nonsuperconductive side represent perspective field for experimental studies.}

\section{4. Experimental verification of existence of localized pairs}

In the preceding sections, two possible types of experiments for detecting of localized pairs in homogeneously disordered materials were singled out. Here we shall review existing experimental data.

 \vspace{2mm}{\bf 4а. Negative magnetoresistance.}
It seems most natural to look for localized pairs in materials which become high resistive after superconductivity is destroyed, i.e. in the vicinity of superconductor-insulator transition. We shall discuss two types of such materials: ultrathin films with thickness $b$ acting as the control parameter (films are superconductors at large $b$ and insulators at small $b$) and materials with composition which can be changed and somehow controlled. In the second group of materials, it is electron concentration and/or disorder level that usually act as control parameters. In both groups, magnetic field $B$ can be used as control parameter.

Let us examine in greater length possible influence of field breaking of localized pairs on the electronic transport. First, the bounding energy has dispersion: it depends on both, specific values of the random potential in the vicinity of the localization point
${\bf r}_i$ and on energy of this electron $\varepsilon^{(1)}$ under switched off superconducting interaction:
\begin{equation} \label{epsilon}
 \varepsilon({\bf r}_i)=\varepsilon^{(1)}({\bf r}_i)-\Delta_L({\bf r}_i) .
\end{equation}

By definition, $\varepsilon({\bf r}_i)<\varepsilon_F$. However, $\varepsilon^{(1)}({\bf r}_i)$ may be both, less and larger than $\varepsilon_F$.
In the first case, $\varepsilon^{(1)}({\bf r}_i)<\varepsilon_F$, the electron remains localized at the point ${\bf r}_i$ after field destroys pair correlations and $\Delta_L$ vanishes. Contribution of this electron into the conductance increases in accordance with eq. (\ref{sigma_2}) though the contribution remains to be hopping. In the second case, $\varepsilon^{(1)}({\bf r}_i)>\varepsilon_F$, the electron delocalizes and his contribution to the conductance becomes metallic.

To ascribe the observed negative magnetoresistance to localization of electron pairs at impurity centers, one must make sure that the material has no granular structure. This was checked in special experiments for all materials discussed below.

\begin{figure}[t]
\includegraphics{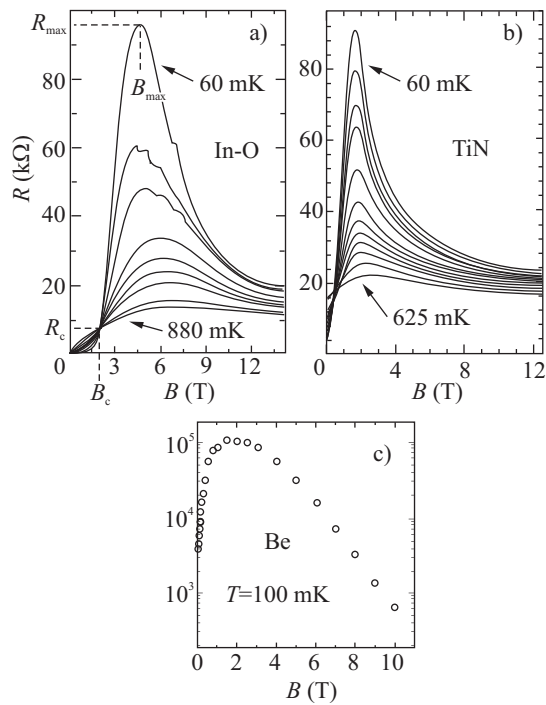}
\caption{Negative magnetoresistance in the normal field in the films made of different materials superconductive without field: (a) In-O film with thickness 20 nm \cite{71-11}; (b) Ti-N film with thickness 5 nm \cite{Batu1}; (c)
film of amorphous Be with thickness slightly below critical, so that the zero field resistance is not zero \cite{WuLT24}} \label{LopaNMr}
\end{figure}

The first experiments where this specific negative magnetoresistance has been observed were performed on amorphous In-O films \cite{PaaHeb2}; it was carefully checked beforehand that granular structure was absent in these films. Advantage of In-O films lies in the comparatively simple way to change the electron concentration and to study the temperature and field dependence of the resistance at different concentrations. Due to this, the negative magnetoresistance on amorphous In-O films was later studied in detail in \cite{71-11,Shahar1}. An example of the field dependence of the resistance $R(B)$ of an \mbox{In-O} film is presented at Fig.~\ref{LopaNMr}(a). The phase transition takes place at the field $B_c$ at which the curves obtained at different temperatures intersect. This field is called critical; at lower fields $B<B_c$, the sample is in superconductive state. The magnitude of resistance $R_c$ in the critical field is of the order of normal resistance of the film.

At field $B$ slightly above $B_c$, the sample belongs to the critical region of the phase transition. Let us omit this region for a while (we shall return to it below, at the end of this section). At higher magnetic field (for the sample which appears on the Fig.~\ref{LopaNMr}(a), one should take field $B\gtrsim B_{\rm max}$), it it possible to define the state of the sample by extrapolation
 $R^{-1}(T)\equiv\sigma(T)\stackrel{T\to\,0}{\longrightarrow}\sigma(0)$.
In the experiment presented on the Fig.~\ref{LopaNMr}(a), it follows from such extrapolation that the sample is insulating in the field range $B_{\rm max}<B<10\,$T and that the insulator region changes into metallic at the field 10~T.

The properties of thus formed insulating state can be characterized by the ratio
$R_{\rm max}/R_c$ at some temperature low enough. For the film at Fig.~\ref{LopaNMr}(a) this ratio is slightly higher than one order of magnitude at 70~mK. Authors of the study \cite{Shahar1} managed to find the state, at which the resistance uprise at 70 mK as compared to the critical $R_c\approx5$~k$\Omega$ was
five orders of magnitude. Temperature dependence of the resistance in wide field interval was of activation type
\begin{equation}\label{expT}
  R(T)=R_0\exp(T_0/T).
\end{equation}
Activation energy $T_0$ depended on magnetic field $B$ and reached maximum about
1.7 K in the field $B_{\rm max}$  \cite{Shahar1}. With further field increase, at $B>B_{\rm max}$, the activation energy gradually declined.

Summarizing, amorphous quasi homogeneous, i.e. not granular, In-O films with oxygen deficit
(i) became insulators when superconductivity was destroyed by magnetic field and remained insulators in wide field diapason;
(ii) with increase of the field, the film resistance reached maximum and then started to decline;
(iii) in the field high enough, transition insulator-metal occurred \cite{71-11} or the sample approached very close to metallic state \cite{Shahar1};
(iv) in the highest field, the film resistance almost returned to the level $R_c$.

All these properties meet the concept of localized pairs with localization owed only to superconducting interaction; the latter is effective due to positive value of electron one-particle energy $\varepsilon^{(1)}$ with respect to  $\varepsilon_F$ (see eq.~(\ref{epsilon})). Negative value of $\varepsilon^{(1)}-\varepsilon_F$ has been also observed in the state of amorphous In-O which was already insulating in zero magnetic field: after applying the field 15~T, the sample retained the activated type of resistance and followed Mott law, i.e. eq.~(\ref{hopping}) with $\nu=4$ \cite{GG}.

The wide range of magnetic fields with negative sign of the derivative $\partial R/\partial B<0$ can be explained by two factors. First, the magnitude $\Delta_L$ reduces with increase of the field gradually:
\begin{equation} \label{DeltaB}
 \Delta_L(B)=\Delta_L(0)-\widetilde g\mu_B|B|
\end{equation}
($\widetilde g$ is effective $g$-factor, and $\mu_B$ is Bohr magneton). Second, the quantity $\Delta_L(0)$ has dispersion by itself. We have mentioned this when considering eq.~(\ref{epsilon}).

A natural question arises: to what extend this phenomenon is universal, i.e. is it a general rule that the superconductor-insulator transition in homogeneous disordered system is accompanied by peak of resistance with negative magnetoresistance  tail in strong magnetic fields. According to Fig.~\ref{LopaNMr}, a similar behavior of the magnetoresistance has been also observed on Ti-N films and ultrathin Be films. Specific electron system at the interface between two layered insulating oxides, LaAlO$_3$ and SrTiO$_3$, behaves similar to Be films \cite{Heterostr}. This two-dimensional system is superconductor at high carrier concentrations and insulator at low concentrations. Normal magnetic field applied to the interface in the insulating state induces increase of the magnetoresistance which is followed by its fall down to the values considerably lower than the starting value at $B=0$.
\begin{figure}
\includegraphics{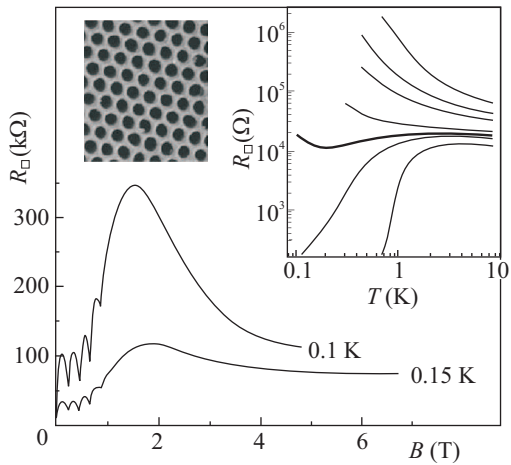}
\caption{Above on the left: aluminum oxide substrate with holes radius $r_{\rm hole}$ about 27 nm and the distance between the hole centers about 100 nm; on this substrate, Bi film was deposited in several small steps which sequentially made it thicker. Above on the right: changes in temperature dependence of the resistance of amorphous Bi film induced by increase of its thickness $b$ (top-down). For the thickness which is singled out by solid line, field dependence of the resistance at two different temperatures is presented below, \cite{NewValles}.} \label{NewValles}
\end{figure}

However, for localized pairs in the insulator to appear, apparently, it is not enough to be close to superconductor-insulator transition. Ultrathin Bi films are a classical example. The very method of experimental studies of superconductor-insulator transitions was developed in experiments with this system, so that it has been  studied in detail \cite{Goldman93}. In standard ultrathin Bi films evaporated with precautions to avoid granular structure, the superconductor-insulator transition is always obtained and the negative magnetoresistance does not exceed one-two percents if exists at all \cite{Valles95,Goldman04,Goldman06}.

In connection with this, amorphous Bi gives additional reason for thoughts. Bi films evaporated in standard way onto perforated substrate demonstrate both, transition {\it and negative magnetoresistance} \cite{NewValles}, see Fig.~\ref{NewValles}.
The curves on the figure contain also oscillations in low fields which are due to so called frustrations. We shall discuss frustrations below, in section 4c. Here they can be considered as indication that a lattice of holes indeed exists in the film bringing forth the special field periodicity.

Summarizing, we assume that
\emph{magnetoresistance peak and negative magnetoresistance in high fields on the insulating side of the superconductor-insulator transition are convincing indications  of pair localization, i.e., of pair  correlations between localized carriers.}  It is difficult to judge resting upon only one experiment \cite{NewValles} whether the localization emerged due to holes in the film or something else. However, there is no doubt that \emph{some factors exist that can enhance or suppress pair correlations of localized carriers in specific material.} We shall return to this question below, in section~5, and at the end of section 4c we shall revisit experiments on perforated substrates.

\vspace{2mm}{\bf 4b. Binding energy of electron pairs --- superconductive
pseudogap.} Let us formulate the definition we shall adhere to. We shall call pseudogap the minimum in the one-particle density of states $g(\varepsilon)$ at the Fermi level which owes to superconductive interaction but exists in the dissipative system. Within this definition come, first, well known for a long time minimum $g(\varepsilon)$ in the fluctuative regime of usual superconductors at $T>T_c$ \cite{VarLark} and, second, electron spectrum in ideal two-dimensional superconductors in zero magnetic field at temperature interval (\ref{BKT}) when Cooper pairs coexist with vortices which provoke dissipation. Finite temperature interval similar to (\ref{BKT}) exists also in magnetic field and in films with disorder.

In essence, influence of localized pairs on the function $g(\varepsilon)$ creates new possibilities for pseudogap to appear. Until last year, there were no experimental studies of $g(\varepsilon)$ and pseudogap in the vicinity of superconductor-insulator transition. Such studies appeared recently based on low-temperature scanning tunnel microscopy. Outstanding promises of this experimental technique and, at the same time, its problems are seen by example of work \cite{SaceTiN-2}, where TiN films were studied.

\begin{figure}[b]
\includegraphics{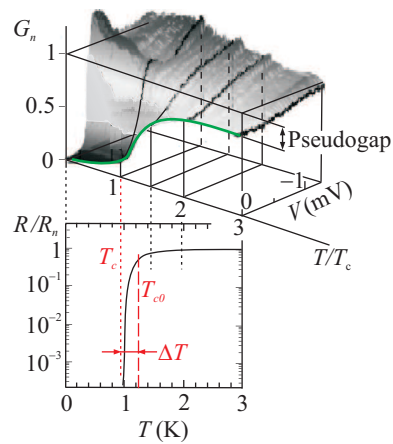}
\caption{Temperature dependence of the density of states at the Fermi level in TiN films ($G_n$ is normalized differential conductivity); solid lines on the surface $G_n(T,V)$ set off the plots $G_n(T)$ at four temperatures, $T_c\approx1$\,K, $1.5\,T_c$, 2\,$T_c$ и 3\,$T_c$, \cite{SaceTiN-2}. Below, the resistive curve of the superconductive transition ($T_c$ is the BKT transition temperature, $T_{c0}$ is the temperature of superconductive transition, $V$ is the bias voltage between the STM tip and the sample, $\Delta T$ is the temperature interval (\ref{BKT}) --- cf.~Fig.~2)}
\label{TiN-2}
\end{figure}

Measurements were done with 5~nm thick TiN films. At each temperature, resistance and VA-characteristics were measured in parallel. This enabled to match evolution of the density of state $g(\varepsilon)$ with the curve of resistive transition.

Fig.~\ref{TiN-2} presents results of such comparison. At the lowest temperatures, the density of states looks as in usual superconductors: dip down to zero in the region $\varepsilon_F\pm\Delta$ with two coherent peaks on both sides. On appearance of dissipation (somewhere near the BKT transition, see Fig.~\ref{Hebard0} for comparison), coherent peaks vanish and the minimum at $\varepsilon_F$ becomes shallow. In this region, only part of electrons are bound into Cooper pairs and this determines the depth of the dip. Pairs move inside the gas of vortices and antivortices which cause fluctuations of the phase of the order parameter; hence, there is no coherence.

Later on, minimum of the function $g(\varepsilon)$ spreads but \emph{survives up to comparatively high temperatures}. The problem is to distinguish whether this minimum indicates existence of localized pairs or it appears due to superconductive interaction in Cooper channel (superconducting fluctuations), or even to Aronov-Altshuler correction \cite{AA2} to  $g(\varepsilon)$ which is caused by interelectron interaction in diffusive channel and hence has nothing to do with superconductivity. This correction is known to increase under grows of disorder turning into Coulomb gap at the metal-insulator transition. Apparently, to separate reliably effect from localized pairs, one would need to combine low-temperature tunnel spectroscopy and strong magnetic field.

\vspace{2mm}{\bf 4c. Size of localized pairs.} There is another manifestation of existence of localized pairs, namely, frustrative oscillations at perforated film in the insulating state \cite{VallesHole}. Ultrathin Bi films were evaporated onto anodized aluminium oxide substrate with lattice of holes of
$r_{\rm hole}=$\,23 nm radius and 95 nm period (see~Fig.~\ref{NewValles}).For conjugating the Bi film with the substrate, a layer of amorphous Ge coated with an
additional Sb layer 1 nm thick was used. For a control, a substrate without holes was placed nearby, onto which the deposition was produced in parallel and which was also tested after each thickening of the Bi films.

Oscillation appearance can be easily explained by the help of notion of frustration $f$, the average number of magnetic flux quanta $\Phi_0$ per array's cell
\begin{equation}\label{frustration}
 f=BS/\Phi_0,\qquad\Phi_0=(2\pi\hbar)/2e,
\end{equation}
where $S$ is the area of the unit cell. Relation (\ref{frustration}) enables to measure magnetic field in units $f$. At integer $f=1,2,...$, the field is concentrated in the holes, so that there is no field in the film; the inhomogeneity of the field is maintained by screening currents. According to classical electrodynamics, this means that persistent current flow around the holes. It follows from the period of oscillation and quantization condition
(\ref{frustration}) that these currents are formed by carriers with charge $2e$, i.e., electron pairs.
\begin{figure}[b]
\includegraphics{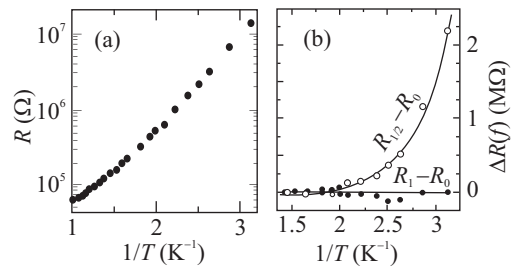}
\caption{Behavior of the perforated Bi film on the non superconductive side of the quantum transition. (а) Temperature dependence of the film resistance which indicates that the film is in the insulating state; (b)~temperature dependence of the magnetoresistance $\Delta R=R_f-R_0$ of the perforated film in the fields $f=1/2$ and $f=1$ compared to the zero-field resistance $R_0$, \cite{VallesHole}}
\label{BiHoles}
\end{figure}

The state chosen for the analysis presented in~Fig.~\ref{BiHoles} is situated slightly deeper in the insulating region than one for which the magnetoresistance was demonstrated in Fig.~\ref{NewValles}. Fig.~\ref{BiHoles}(a) shows that the resistance in this state grows exponentially with temperature decrease with activation energy about 5~К. At the same time, there are frustrative oscillations on the film in this state. They are characterized on the Fig.~\ref{BiHoles}(b): the curve with empty circles demonstrates, how the amplitude of the oscillations grows the temperature decrease; the curve with full circles argues that there is no monotonous contribution in the region of first oscillations. Hence, it follows from Fig.~\ref{BiHoles}(b) that on the scale of $\sqrt{S}$ superconducting currents do exist, whereas on the scale of the sample dimension, according to Fig.~\ref{BiHoles}(a), there are neither superconducting currents nor zero-temperature conductivity at all.

Let us assume that the lattice of holes did not affected quasi homogeneity of the film near the superconductor-insulator transition and that, close to the transition at the insulating side, there are localized pairs with localization length $\Lambda_2$ which satisfy inequality (\ref{Lambda_2}). This enables to interpret the experiment \cite{VallesHole} as measurement of the lower limit for the length of pair localization $\Lambda_2$
\begin{equation}\label{summ1}
  r_{\rm hole}<\Lambda_2.
\end{equation}
\emph{for the specific film presented on the Fig.~\ref{BiHoles} at the specific values of the control parameters.}

In accordance with this interpretation, frustrative oscillations can be observed only near the transition; they are absent on films with higher resistivity which are situated deeper in the insulating region. So far, there is no theoretical interpretation of such "local Meissner effect"\ in the macroscopic insulator. In particular, it is not clear what additional constraints for observation of this effect may arise from superconducting penetration depth.

Let us revisit now the positive magnetoresistance in In-O (Fig.~\ref{LopaNMr}a) at the left slope of the magnetoresistance peak in the field interval  \begin{equation}\label{B-B}
  B_c>B>B_{\rm max}.
\end{equation}
It can be qualitatively explained by gradual decrease of the length $\Lambda_2$ with the state shifting away from the transition inside the insulating region.
The conductance in the field interval (\ref{B-B}) is assumed to be determined by diffusion and hopping of localized pairs. Hence, decrease of $\Lambda_2$ with  field growth in this interval is accompanied by lowering of hopping probability and increase of the resistance. However, the field affects $\Lambda_2$ simultaneously in the opposite way: the growth of the field decreases the binding energy and increases $a_{2B}$ and, as a result,  $\Lambda_2$. Two opposite effects may lead to widening of the interval (\ref{B-B}); its right edge $B_{\rm max}$ is determined by condition $\Lambda_2\approx a_{2B}$ when the first effect becomes unimportant.

\section{5. Additional factors that promote pair localization}

In the simplest case, two electrons presenting a pair may be localized in one large well of the random potential. Such configuration differ from a small grain only in lack of a high barrier along its perimeter. Essential limitation for such pairing to occur comes from the volume occupied by a localized electron. Formula (\ref{b1b2}) establishes the lower bound $b_2$ for the size of isolated grain. At $b>b_2$, superconductive interaction does not affect the electron spectrum inside the grain.  For an electron localized on a defect, localization length $\Lambda$ of its wave function figures as the size $b$ and represents the first important factor for stimulation of superconductive correlations between localized electrons.

It is not necessary for the electrons forming a pair to be localized on the same defect: the phonon-based attraction is long-range. However,  the mean distance between the electrons $s\sim (g_0\Delta)^{-1/3}$ which we have estimated in section 3b in connection with inequality (\ref{1}) may prove to be too large. For pairing, it is important that the distance between the localization centers were as short as possible. So, availability of closely disposed centers is the second important factor which stimulates pairing. Each of localized electrons should at the same time occupy volume large enough.

Let us consider these two factors.

\begin{figure}[b]
\includegraphics{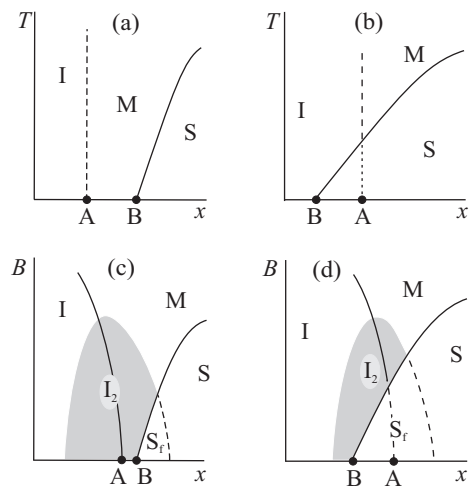}
\caption{ (a) and (b): two versions of the phase diagram
insulator-metal-superconductor (I-M-S) on the plane $(x,T)$ in zero magnetic field; (c) and (d): two corresponding versions of the phase diagram on the plane $(x,B)$ at $T=0$. Regions I$_2$, where fractal wave functions stimulate superconductive interaction between localized electrons, are marked out by grey.  S$_f$ are expected regions of fractal superconductivity}
 \label{diagrams}
\end{figure}

\vspace{2mm}{\bf 5a. Proximity of metal--insulator transition, either real or
virtual.}
Generally speaking, on change of the control parameter $x$, for instance, after increase of electron concentration, a disordered insulator may turn either into normal metal, or into superconductor. The full phase diagram contains not two, but three states of the electron system: insulator (I), normal metal (M), and superconductor (S). Rough draft of two alternatives of the phase diagram on the plane $(x,T)$ for a three-dimensional electron system is presented on Fig.~\ref{diagrams}(a),(b). On diagram (a), two transitions, I$\to$M$\to$S, occur in series. As is known, the metal-insulator transition is imaged as a point on the $x$-axis of the plane $(x,T)$ (see, for instance, \cite{GD1}). Let it be point $A$. The vertical dashed line from the point $x=A$ at Fig.~\ref{diagrams}a does not mark a real phase boundary. It indicates that extrapolation of the conductivity to $T=0$ will give zero in the strip I but finite value in the strip M. To the right from point $B$ the zero-temperature conductivity is infinite.

When the transition in the system is single-stage: I$\to$S, then virtual point $x=A$ in the superconductive region exists all the same. If it were possible to switch off the superconducting interaction, the system would not remain insulator but would turn into normal metal at the value $x{=}A{>}B$ of the control parameter. This variant is presented on Fig.~\ref{diagrams}(b).

Points A and B usually are situated close to each other, so that to determine which variant exists in specific system, serious experimental efforts should be used. Localized pairs are expected to appear in the vicinity of point B and proximity to point A affects the probability of pairing processes. This idea was formulated first in \cite{FeigKravts}; afterwards, the model was developed in detail in the next work of these authors \cite{FeigKravts2}. The mechanism of the influence may be described as follows.

Far from the metal-insulator transition, the localization length $\Lambda$ which enters the asymptotic (\ref{aB}) is determined by Bohr radius, $\Lambda\approx a_B$. While approaching of the transition, the tails of the wave functions begin to overlap and the localization length $\Lambda$ to grow. Law of $\Lambda$ variation on approaching the transition depends on the structure of the wave functions. In the vicinity of the transition, on the scales lower then $\Lambda$, wave function have fractal structure with fractal dimension $D_f<3$.
According to numerical calculations \cite{Mirlin}, in the vicinity of a standard 3D Anderson transition, $D_f=1.30\pm0.05$. Fractality of the wave function makes it more friable and increases its typical size $\Lambda$. Special calculations made in \cite{FeigKravts2} found that Cooper pairing could be successfully realized on fractal wave functions.

In terms of phase diagrams (a) and (b) from Fig.~\ref{diagrams}, one may say that to the left from point A an interval of control parameter values exists where the wave functions of localized electrons swell due to fractality and hence become liable to superconductive interaction. Localization length $\Lambda$ in this interval fits the inequality
\begin{equation}\label{LambdaD}
  \Lambda_D\leqslant\Lambda\leqslant\infty.
\end{equation}
It is limited from below by the condition $\delta\varepsilon=\delta\varepsilon(\Lambda_D)=\hbar\omega_D$.

The electron wave functions are fractal on both sides from Anderson transition. Length which diverges at the transition exists on the metallic side too; usually it is named correlation length and designated as $\xi$. When moving away from the transition, the length $\xi$ gradually lessens and transforms into the mean free path $l$ (cf. with inequalities (\ref{Lambda}) and  (\ref{Lambda_2})):
\begin{equation}\label{xi}
  l\leqslant\xi\leqslant\infty.
\end{equation}
Wave functions for the set of states with length $\xi$ from the interval (\ref{xi}) behave at large distances $r\gg\xi$ in a usual way typical for delocalized electrons.

Summarizing, \emph{proximity to metal-insulator transition promotes forming of localized pairs in the vicinity of the superconductor-insulator transition.}

Regions on the plane $(x,B)$ where fractality facilitates pairing of localized electrons are labeled  $I_2$ and colored grey on Fig.~\ref{diagrams} (c) and (d). These regions are bounded above processes of pair breaking by magnetic field. It is situated mainly to the left from point B on the diagram (c) and mainly to the right on the diagram (d). Basing on experimental negative magnetoresistance data, it seems that both variants really exist: variant (c) in Be films and in the heterostructure LaAlO$_3$/SrTiO$_3$, variant (d) in InO and TiN films (this follows also from other experimental data collected in review \cite{GD2}).

On Fig.~\ref{diagrams} (c) and (d), extremely interesting superconductive regions S$_f$ are marked out. In works \cite{FeigKravts} and \cite{FeigKravts2}, they are named regions with fractal superconductivity. Apparently, study of fractal superconductivity is a future task.

  \vspace{2mm}{\bf 5b. "Chemical predisposition"\ to pair localization.}
Correlations in the random potential are known to change drastically localization properties of the system. This exhibits especially sharply in one-dimensional case. For instance, if impurities are located in pairs with fixed distance between them, then in spite of random disposition of these pairs, electrons with specific energy values turn to become delocalized (dimer model \cite{Dimer}).

\begin{figure}[h]
\includegraphics{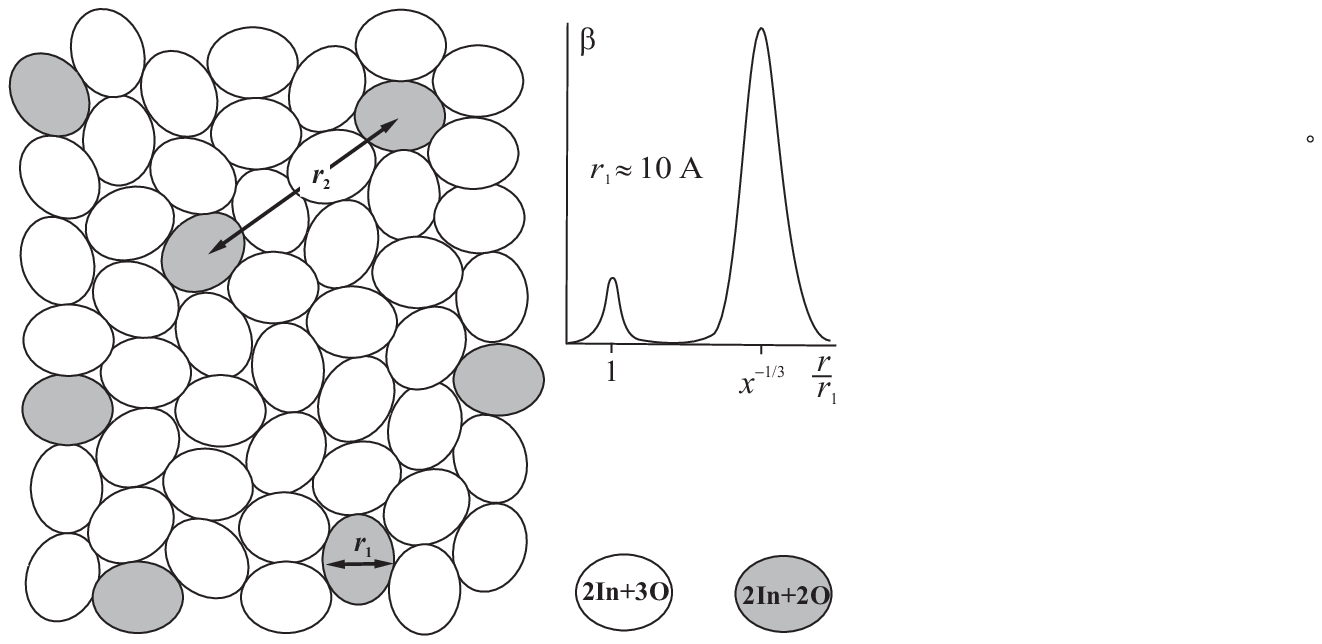}
\caption{Conditional view for distribution of oxygen vacancies in amorphous
$\rm In_2O_{3-\it x}$. Two electrons may localize in each grey region. The function $\beta(r)$ on the plot is probability for one electron to be localized at the distance $r$ from the other}
 \label{CorrRand}
\end{figure}

Apparently, correlations of the random potential may turn to be important for the process of localized electron pairing too. Let us consider amorphous In-O as an example. Configuration of five atoms, [2In+3O], is a structure unit of this material. All valence electrons are utilized in covalent bonds inside this unit, just as in molecule In$_2$O$_{3}$; hence they are effectively bound.
Chemical content of the real amorphous material is described by formula
In$_2$O$_{3-x}$. Portion $x$ of structure units have oxygen vacancy and two valence electrons in the vicinity of each vacancy are weakly bound to ion skeleton. They easily become delocalized leaving pairwise correlated wells in the random potential (see Fig.~\ref{CorrRand}).

Let us introduce probability $\beta=n(Q(r)+1)d^3r$ for an electron to be in the volume $d^3r$ given that another electron is localized in the origin $r=0$. Here $Q(r)$ is pairwise correlation function (if positions of all localized electrons are statistically independent, $Q(r)\equiv0$); $n$ is concentration of electrons which are not utilized in covalent bonds and can be delocalized as well as localized. It can be assumed for estimate that $n=2xV^{-1}$, where $V$ is average volume of one structure unit approximately equal to the unit cell volume $V_0$ of the crystal In$_2$O$_{3}$, $V\approx V_0\approx10^3$\AA$^3$, so that
$n\approx2x\times10^{21}$cm$^{-3}$.

Freehand sketch of the function $\beta(r)$ is presented in the upper right corner of Fig.~\ref{CorrRand}. Position of the first maximum is determined by the average size of the structure unit, $r_1\approx10\,$\AA, position of the second by concentration of oxygen vacancies, $r_2\approx r_1x^{-1/3}$. Maximum at comparatively small $r_1<n^{-1/3}$ causes the "predisposition"\ noted in the title of the section.

\vspace{5mm}
{\bf Acknowledgement} Author thanks V.T. Dolgopolov for valuable
discussions. This work was supported by contract 02.740.11.0216 from  the RF Ministry of Education and Science.

\end{document}